\documentclass[aps,prl,twocolumn,showpacs]{revtex4}
\usepackage{graphicx}
\usepackage{amssymb}

\begin{document}

\title{New approximation for nonlinear evolution in periodic potentials}

\author{Micha\l{} Matuszewski}

\affiliation{Institute for Theoretical Physics, Warsaw
University, Ho\.{z}a 69, PL-00-681 Warsaw, Poland}

\begin{abstract}
A new approximation for evolution described by Nonlinear 
Schr\"odinger Equation (NLS) with periodic potential is presented. 
It relies on restricting dynamics to one band of the bandgap spectrum,
and taking into account only one, dominating Fourier component in the nonlinear Bloch-wave mixing.
The resulting equation has a
simple, discrete form in the basis of linear Wannier functions, and turns out to be very
accurate as long as the modes in other bands are not excited and the potential 
is not very deep. Widely used approximations, 
the tight-binding approximation and the effective mass approximation,
are derived from the new equation as the limiting cases.
\end{abstract}

\pacs{03.75.Lm, 05.45.Yv, 42.65.Tg}

\maketitle

The Nonlinear Schr\"odinger Equation (NLS) is one of the
most common and basic nonlinear wave equations.
Recent experimental progress in the fields of nonlinear
optics 
and Bose-Einstein condensation, where it is known as
the Gross-Pitaevskii equation, has strongly stimulated theoretical
studies on its properties and solutions. 
The ability to describe and understand nonlinear evolution in periodic potentials
is of a great practical importance, since this kind of potential 
can be easily created by a periodically varying refractive
index of an optical medium~\cite{KivsharAgrawal,Eisenberg,Fleischer,SilberbergNature}, 
or by a standing laser wave imposed
on a Bose-Einstein condensate~\cite{OberthalerRMP,Oberthaler}. On the other hand, 
the presence of a periodic potential
can dramatically modify the wave diffraction, and creates a
way to control the nonlinear evolution of physical systems.

To date, there are a few models commonly used to
describe these systems in a simplified way. 
In the case of a deep potential, the essential properties of 
system dynamics are included in the tight-binding (or coupled-mode)
approximation, which assumes that the interaction of neighbouring cells
can be described by tunneling between their localized modes.
As a result of this assumption, one obtains a Discrete Nonlinear 
Schr\"odinger Equation (DNLS), a very convenient tool for studying and 
modelling e.g.~light propagation in weakly coupled waveguide arrays~\cite{KivsharAgrawal,Eisenberg,SilberbergNature}. 
If the Bloch-wave spectrum of the system contains only a small part of the Brillouin zone,
one can use the effective mass approximation to describe 
motion of a wavepacket spanning over many sites of the potential~\cite{OberthalerRMP}.
It was successfuly used to model the creation of gap solitons --
nonlinear stationary states exisiting even in the case of
a negative (self-defocusing, repulsive) nonlinearity~\cite{Oberthaler,GAPBEC}.

In this letter, a new approximation is presented, which brings together
positive aspects of these two models, while preserving most of their simplicity. The derivation relies on assuming that 
the dynamics restrict to one band, and taking into account only one, dominating Fourier component 
in the nonlinear Bloch-wave mixing. It leads to a
simple discrete equation in the basis of linear Wannier functions~\cite{Alfimov,Bhraznyi}.
The new equation can be used to describe evolution in weak or deep potentials (as long as they are not very deep),
and includes dynamics in the whole Brillouin zone. Its accuracy is better than could be expected,
as will follow from direct numerical comparison of its results to solutions of the NLS equation.

In reduced units, the Nonlinear Schr\"odinger Equation with periodic potential 
for time-dependent function $u(x,t)$ reads
\begin{equation} 
i \frac{\partial u}{\partial t} = \frac{\partial^2 u}{\partial x^2} - U(x) u +  \sigma |u|^2 u\,,
\label{NLS}
\end {equation}
where $U(x+1)=U(x) \in \mathbb{R}$ and $\sigma= \pm 1$ is the sign of the nonlinearity. 
The norm of the solution $N=\int_{-\infty}^{+\infty}|u|^2 \mathrm{d}x$ is a constant of motion.
After performing Fourier transform one obtains the equation for 
$\tilde{u}(k,t)=(2 \pi)^{-1/2} \int_{-\infty}^{+\infty} u(x,t) \mathrm{e}^{-ikx} \mathrm{d}x$
\begin{eqnarray} \label{FNLS}
i \frac{\partial \tilde{u}(k)}{\partial t} &=& -k^2 \tilde{u}(k) 
- \sum_{n} U_n \tilde{u}(k - 2 \pi n) +\\ \nonumber
&+& \frac{\sigma}{2 \pi}  \int\!\!\!\int_{-\infty}^{+\infty} u(k_1) u^*(k_2) u(k-k_1+k_2)  \mathrm{d}k_1 \mathrm{d}k_2\,,
\end{eqnarray}
where $U(x)=\sum_{n} U_n \mathrm{e}^{2 i \pi n x}$, and the integer index $n \in (-\infty, +\infty)$.

From now on the dynamics will be restricted to one certain band of the bandgap spectrum.
The pseudomomentum will be denoted with $k'$.
For each $k' \in (-\pi,\pi)$, there exists a Bloch function $\nu(x,k')$ of energy $E(k')$, 
and its Fourier transform $\nu(k,k')$ is the solution of the linear version of Eq.~(\ref{FNLS})
\begin{equation} \label{blochdef}
E(k') \nu(k, k')  = k^2 \nu(k, k') + \sum_n U_n \nu(k - 2 \pi n, k'),
\end{equation}
with orthonormality condition $\int \nu(k,k'_1) \nu^{*}(k,k'_2) {\rm d}k =
\delta(k'_1-k'_2)$. 
It is assumed that the solution is  
the superposition of Bloch functions of the chosen band
\begin{equation} \label{decompose}
\tilde{u}(k, t)  = \int_{-\pi}^{\pi} \tilde{\phi}(k',t) \nu(k,k') {\rm d}k'\,.
\end{equation}
This assumption implies that the nonlinearity is rather weak, since strong nonlinearity
leads to Bloch wave mixing of waves in different bands, or even destruction of the band structure~\cite{OberthalerRMP}.
After substituting this equation into (\ref{FNLS}) and using (\ref{blochdef}), one can multiply both sides by $u^*(k,k_0')$ and
integrate over $k$ to obtain
\begin{eqnarray*}
&i&\!\!\!\! \frac{\partial \tilde{\phi}(k_0', t)}{\partial t} = -E(k_0') \tilde{\phi}(k_0', t) + \\
&+&\frac{\sigma}{2\pi} \int \!\!\!\int \!\!\!\int_{-\pi}^{\pi}
\tilde{\phi}(k'_1, t) \tilde{\phi}^{*}(k'_2, t) \tilde{\phi}(k'_3, t) \mathrm{d}k'_1 \mathrm{d}k'_2 \mathrm{d}k'_3 \times  \\
&\times& \int \!\!\!\int \!\!\!\int_{-\infty}^{+\infty}
\nu^{*}(k_0,k_0') \nu(k_1,k'_1) \nu^{*}(k_2,k'_2) \times \\
& \times& \nu(k_0 - k_1 + k_2,k'_3) \mathrm{d}k_0 \mathrm{d}k_1 \mathrm{d}k_2 .
\end{eqnarray*}
\begin{figure}[tbp]
\includegraphics[width=8.5cm]{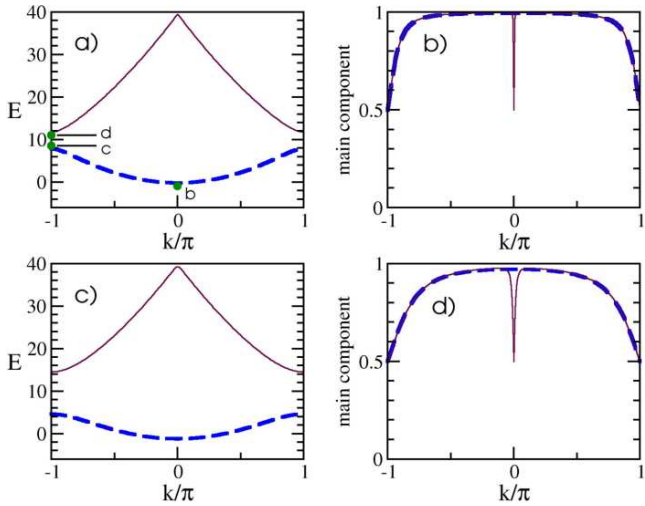}
\caption{
(color online) a) Linear Bloch wave spectrum in cosine potential of depth $\varepsilon=4$. 
Small dots correspond to frames in Fig.~\ref{profiles}.
b) Contribution of the main component in decomposition of a Bloch wave in Fourier space.
Dashed and solid lines correspond to the first and the second band, respectively.
c), d) Same as a), b), but for potential depth $\varepsilon=10$.
}
\label{bloch}
\end{figure}
For this equation to be useful, the nonlinear term has to be simplified.
According to Bloch theorem, Fourier spectrum of a Bloch function is discrete,
$\nu(k,k')=\sum_n \nu_n(k') \delta(k-k'-2\pi n)$. 
It turns out that in most cases one of the Fourier components dominates, as can be 
seen in Fig.~\ref{bloch}, where the results for cosine potential $U(x)=-\varepsilon \cos(2\pi x)$ are shown.
Let $n_{\rm band}$ be the band number, starting from zero for the lowest one;
then one can assume that $\nu(k,k') \approx \delta(k-k_{\rm main})$, where 
$k_{\rm main}= k' + {\rm signum}(k') \pi n_{\rm band}$ for even $n_{\rm band}$,
and $k_{\rm main}= k' - {\rm signum}(k') \pi (n_{\rm band} + 1)$ for odd $n_{\rm band}$.
This property is related
to the fact that in the limit of vanishing potential Bloch decomposition is equivalent to the Fourier transform.
From Fig.~\ref{bloch} one can see that this assumption is fulfilled better for a weaker potential, and 
that the agreement is worst at the edges of the Brillouin zone and in its center for higher bands 
(these are the points of jump in $k_{\rm main}$). However, as it will be shown below, this approximation
gives good results even in the case of gap solitons, nonlinear objects which reside in the vicinity
of these points.

As a result, the above equation takes the form
\begin{eqnarray*}
&i&\!\!\!\! \frac{\partial \tilde{\phi}(k', t)}{\partial t} = -E(k') \tilde{\phi}(k', t) + \\
&+&\frac{\sigma}{2\pi} \int  \!\!\!\int_{-\pi}^{\pi}
\tilde{\phi}(k'_1, t) \tilde{\phi}^{*}(k'_2, t) \tilde{\phi}(k'-k_1'+k_2', t) \mathrm{d}k'_1 \mathrm{d}k'_2.
\end{eqnarray*}
Inverse Fourier transform gives a simple discrete equation, which is the main result of
this letter
\begin{equation} \label{new}
i \frac{\mathrm{d} \phi_n}{\mathrm{d} t} = - \sum_m E_m \phi_{n-m} + \sigma |\phi_n|^2 \phi_n.
\end{equation}
Here $\tilde{\phi}(k',t)=(2 \pi)^{-1/2} \sum_n \phi_n(t) \mathrm{e}^{-ik'n}$ and
$E(k')=\sum_n E_n \mathrm{e}^{-ik'n}$, $E_{-n}=E^*_n$. The norm of the discrete function is $\sum_n |\phi_n|^2=N$.
Most of the terms in the infinite sum can be neglected, since absolute values of $E_m$ are
significant only for $m$ close to zero. In numerical simulations presented below,
only terms with $|m| \leq 5$ were taken into account. Values of $E_m$ can be found quite easily,
by solving a linear eigenvalue problem (\ref{blochdef}) (see e.g.~\cite{OberthalerRMP}) and performing an inverse Fourier transform.

\begin{figure}[tbp]
\includegraphics[width=8.5cm]{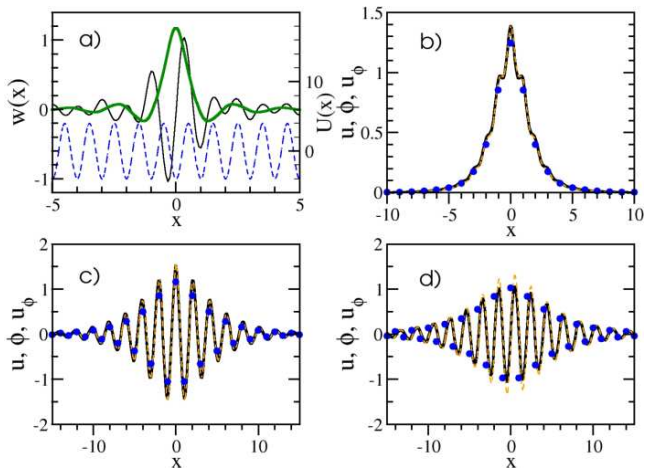}
\caption{
(color online) 
a) Localized Wannier function of the first band (thick line) and the second band
(thin line) for cosine potential depth $\varepsilon=4$. 
Dashed line shows the shape of the potential.
b) Stationary soliton state at the bottom of the first band (see Fig.~\ref{bloch}a), according
to the full NLS model (Eq.~\ref{NLS}, solid line), and the approximate model
(Eq.~\ref{uap}, dashed line, the two lines overlap). Dots show the amplitude
of the discrete function $\phi_n$. Parameters are: $\varepsilon=4$, $\sigma=1$, $N=3.4$
and $\beta=-1$.
c) Soliton at the top of the first band for $\sigma=-1$, $N=7$
and $\beta=8.5$. d) Soliton at the bottom of the second band for $\sigma=1$, $N=6.9$
and $\beta=11$. 
}
\label{profiles}
\end{figure}

The scheme of the above derivation is similar to the one presented in \cite{Alfimov}, where a vector
discrete equation was obtained for evolution in the basis of Wannier functions.
In fact, the one-band assumption imply that the function $u(x,t)$ can be approximated by
\begin{equation} \label{uap}
u(x,t) \approx u_\phi(x,t) = \sum_n \phi_n(t)\, w(x-n),
\end{equation}
where $w(x)$ is the linear Wannier function $w(x)=\int_{-\pi}^{\pi} \nu(x, k')\mathrm{d}k'$, see Fig.~\ref{profiles}a).
Therefore, $\phi_n$ can be interpreted as the amplitude of the wavefunction in the $n$-th
site of the periodic potential.

It is relevant to point out possible generalizations of the presented model. 
The one-band approximation can be extended to include
more than one band. In this case, one would obtain a system of
discrete equations \cite{Alfimov}, each of them coupled with others by 
cross-phase modulation terms. 
The derivation can also be easily generalized to the case of multi-dimensional NLS
equation \cite{Fleischer,Baizakov}. However, one has to be aware that in this case the nonlinear mixing 
between bands is more likely to occur, since the band gaps are not always closed \cite{OpenGap}. 
Finally, the presented method can be applied to other types of nonlinearity,
e.~g.~quadratic, cubic-quintic, or other nonlinearities after
expanding them in Taylor series. A detailed study on these generalizations
will be presented elsewhere.

Equation (\ref{new}) has two interesting limiting cases. For a deep potential,
the energy dependence $E(k')$ becomes close to the cosinus function, 
and the infinite sum can be approximated by three terms with $m=-1,0,1$. This form 
of the equation is equivalent to the one obtained within the tight-binding approximation 
\cite{KivsharAgrawal,Eisenberg,OberthalerRMP}.
However, the new equation will not be accurate, since for deep potentials
the assumption made in the derivation is not well satisfied, see Fig.~\ref{bloch}d).
Still, it will give qualitatively correct results.

Another limit corresponds to the case when the function $\phi$ can be written in the form
$\phi_n(t)=\Phi(n,t) \exp(i k'_0n)$, where $\Phi(x,t)$ is an envelope 
slowly varying in the spatial coordinate $x$. Using this relation and 
expanding $\Phi(x,t)$ in Taylor series up to the second order leads to the 
effective mass equation
\begin{equation} \label{meff}
i\frac{\partial \Phi}{\partial t} = -E(k'_0) \Phi - i v_{\rm g} \frac{\partial \Phi}{\partial x}
+ \frac{1}{m_{\rm eff}} \frac{\partial^2 \Phi}{\partial x^2} + \sigma |\Phi|^2\Phi\,,
\end{equation}
where $v_{\rm g} = (\mathrm{d} E / \mathrm{d} k')|_{k'=k'_0}$ is the group velocity, and 
$m_{\rm eff} = (\mathrm{d}^2 E / \mathrm{d} k'^2)^{-1}|_{k'=k'_0}$
is the effective mass. In this case, the new equation (\ref{new}) agrees
{\it quantitatively} with the NLS equation (\ref{NLS}). This is confirmed by numercial simulations, 
see Fig.~\ref{profiles}.

To test the new approximation, the equation (\ref{new}) was applied
to description of band-gap solitons~\cite{KivsharAgrawal,OberthalerRMP}. 
These states are the inherently nonlinear solutions of the NLS equation (\ref{NLS}) in the form
$u(x,t)=u(x) \exp(-i \beta t)$, where $\beta$ is the eigenvalue of the soliton,
lying in the gap of the band-gap spectrum. In Fig.~\ref{profiles} these states are compared with analogous solutions
of Eq.~(\ref{new}), $\phi_n(t)=\phi_n \exp(-i \beta t)$ and the corresponding functions $u_{\phi}$, 
cf.~Eq.~(\ref{uap}). In particular, the two figures a), b) present solutions of the same equation,
describing evolution in the lowest band, with the only change
in the sign of the nonlinearity. The agreement with the full NLS solution is perfect.
Interestingly, the solution in Fig.~\ref{profiles}b) has larger width and norm
than the solution in Fig.~\ref{profiles}a). In the tight-binding model,
solitons with larger norm always has a smaller width, independently on the sign of nonlinearity~\cite{KivsharOL}.
Here, the new equation takes into account the difference in diffraction strength at the top and the bottom of the first band.
In Fig.~\ref{profiles}c) a soliton composed of Bloch waves from the second band is shown. In this case,
maxima of amplitude lie on maxima of the periodic potential. The agreement between the full model and the approximate
model is in this case somewhat inferior. Additional simulations
have shown that the reason for this is the strength of the nonlinearity; the eigenvalue of the solution
is shifted deeper into the band gap than in the previous cases. In general, the equation (\ref{new})
works best if the nonlinear energy is much smaller than both the gap width and the band width.

\begin{figure}[tbp]
\includegraphics[width=8.5cm]{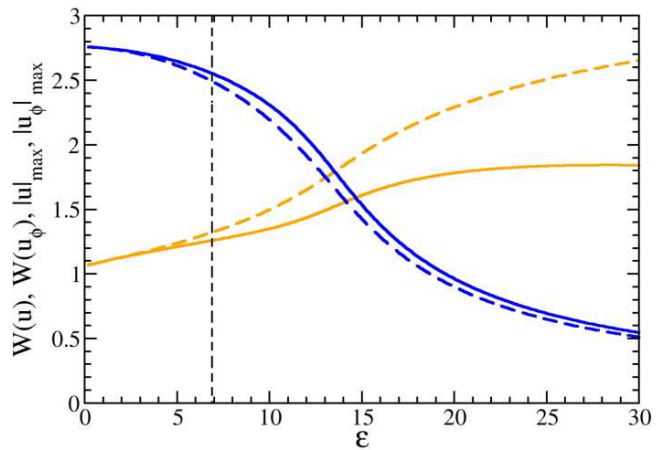}
\caption{
(color online) 
Comparison of the calculated soliton width (dark lines) and maximum amplitude (light lines)
according to the NLS equation (\ref{NLS}) (solid) and the approximation (\ref{uap}) (dashed),
versus the potential depth $\varepsilon$. Structure of solitons is the same as in in Fig.~\ref{profiles}b). 
For each $\varepsilon$, the norm of the NLS solution is $N=3$, and
the eigenvalues $\beta$ of the two compared solutions are equal.
The vertical line indicates the value of $\varepsilon$
when the widths of the first band and the first gap are the same. This is the
point of transition from the weak potential regime to the deep potential regime.
}
\label{varU}
\end{figure}

In Fig.~\ref{varU} a systematic comparison of the full and approximate models is presented.
Here the width and maximum amplitude of the lowest-energy soliton is depicted versus the potential depth $\varepsilon$.
Good agreement of calculated width $W(u)=3 \int |x| |u(x)|^2 \mathrm{d} x \,/ \int |u(x)|^2 \mathrm{d} x$
suggests that the shape of the approximate solution is correct even for very strong potentials, 
with the only difference in the norm $N$.

\begin{figure}[tbp]
\includegraphics[width=8.5cm]{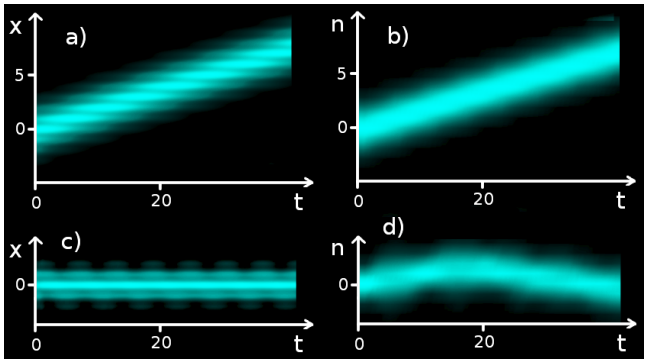}
\caption{
(color online)
Motion of a soliton wavepacket with imprinted linear phase for $N=3$ and $\varepsilon=4$,
according to a) the NLS equation and b) the discrete equation (\ref{new}).
c), d) Same as in a), b), but for $\varepsilon=10$. In this case, the soliton is trapped 
by the Peierls-Nabarro potential. The shape of solitons is the same as in Fig.~\ref{profiles}b).
}
\label{pn}
\end{figure}

The new approximation can also desribe dynamical evolution problems. 
A very simplified version of it was already used for justification of the phenomenon
of spontaneous migration of Bloch waves to the regions of normal diffraction
for positive nonlinearity, and to the regions of anomalous diffraction for
negative one~\cite{GAPBEC}. This nolinear phenomenon was observed experimentally
in two different systems~\cite{OE}, and can be utilized for efficient soliton generation~\cite{GAPBEC}. 
Here the new model is used
to describe soliton mobility, see Fig.~\ref{pn}.
The lowest-energy soliton wavepacket was ``boosted'' by imprinting a linear phase $u(x,0)=u(x) \exp(i k x)$,
with $k=0.1$. In the case of a weak potential, the wavepacket started to move
with a constant velocity. In the case of a deep potential, it has been trapped
in central sites by the Peierls-Nabarro potential~\cite{PN}.
This effect would not be seen within the usual effective mass approximation.
Here, the agreement between the NLS equation and Eq.~(\ref{new}) is very good for the weak potential,
worse for the deep potential, but the main effect is still apparent. 

In conclusion, a new approximation for evolution described by Nonlinear 
Schr\"odinger Equation with periodic potential was presented. 
The derivation is based on the one-band approximation and simplification
of nonlinear Bloch-wave mixing, and leads to a
simple discrete equation in the basis of linear Wannier functions.
The equation works very good as long as the nonlinearity
does not cause excitation of modes in other bands and the potential is not very deep. 
The new model was used for description of gap solitons and a dynamical evolution.
It was shown that the tight-binding approximation and the effective mass approximation
can be derived from the new equation as the limiting cases.
Possible generalizations were pointed out.

The author acknowledges support from the Foundation for Polish Science.

\end{document}